\documentclass[preprint2]{aastex62}
\usepackage{natbib}
\accepted{July 23, 2019}
\shorttitle{Shock Heating Energy in Umbral Flashes}
\shortauthors{Anan et al.}
\begin{document}

\title{Shock Heating Energy of Umbral Flashes \\ Measured with Integral Field Unit Spectroscopy}

\correspondingauthor{Tetsu Anan}
\email{tanan@nso.edu}

\author[0000-0001-6824-1108]{Tetsu Anan}
\affil{National Solar Observatory \\
22 Ohi`a Ku, Makawao \\
Hawaii, 96768, USA}
\nocollaboration

\author[0000-0002-7451-9804]{Thomas A. Schad}
\affil{National Solar Observatory \\
22 Ohi`a Ku, Makawao \\
Hawaii, 96768, USA}
\nocollaboration

\author[0000-0001-5459-2628]{Sarah A. Jaeggli}
\affil{National Solar Observatory \\
22 Ohi`a Ku, Makawao \\
Hawaii, 96768, USA}
\nocollaboration

\author[0000-0002-8259-8303]{Lucas A. Tarr}
\affil{George Mason University \\
Fairfax, Virginia, 22030, USA}



\begin{abstract}

Umbral flashes are periodic brightness increases routinely observed in the core of chromospheric lines within sunspot umbrae and are attributed to propagating shock fronts.
In this work we quantify the shock heating energy of these umbral flashes using observations in the near infrared \ion{He}{1} triplet obtained on 2014 December 7 with the SpectroPolarimetric Imager for the Energetic Sun (SPIES), which is a novel integral field unit spectrograph at the Dunn Solar Telescope.
We determine the shock properties (the Mach number and the propagation speed) by fitting the measured \ion{He}{1} spectral profiles with a theoretical radiative transfer model consisting of two constant property atmospheric slabs whose temperatures and macroscopic velocities are constrained by the Rankine-Hugoniot relations.
From the Mach number, the shock heating energy per unit mass of plasma is derived to be $2 \times10^{10}$ erg ${\rm g^{-1}}$, which is insufficient to maintain the umbral chromosphere.
In addition, we find that the shocks propagate upward with the sound speed and the Mach number does not depend on the temperature upstream of the shocks.
The latter may imply suppression of the amplification of the Mach number due to energy loss of the shocks.
\end{abstract}



%%%%%%%%%%%%%%%%%%%%%%%%%%%%%%%%%%%%%%%%%%%%%
%%%%%%%%%%%%%%%%%%%%%%%%%%%%%%%%%%%%%%%%%%%%%
\section{Introduction} \label{sec:intro}

In the solar chromosphere, the temperature increases with height and the radiative emission is larger than that expected from radiative equilibrium \citep{vernazza81, maltby86}.
Many kinds of heating mechanisms have been suggested to balance the large excess of radiative cooling energy, for example, shock heating \citep{schwarzschild48, carlsson97, beck08}, magnetic energy dissipation in current sheets \citep{parker83, solanki03, hector05, tritschler08}, viscous and Ohmic dissipations in vortical magnetic structures \citep{moll12}, Alfv${\rm \acute{e}}$n wave turbulence \citep{ballegooijen11}, and dissipation resulting from collisions between magnetized ions and unmagnetized neutral atoms \citep{osterbrock61, khomenko12}.
Among the challenges to distinguish which heating mechanisms dominate is establishing remote sensing techniques to estimate each mechanism's heating energy. 

Evidence that shocks play a role in chromospheric heating comes from umbral flashes.
The umbral flashes are a ubiquitous feature of the dynamic chromosphere above sunspot umbrae.
They manifest as periodic brightness increases in the core of chromospheric spectral lines with a period of $\sim 3$ minutes \citep{beckers69, wittmann69, khomenko15}.
Temporal Doppler velocity fluctuations are correlated with the umbral flash \citep{kneer81}.
Using spectroscopic observations, \citet{giovanelli78} interpreted the velocity fluctuations as upward propagating waves, and \citet{lites84} found that the propagating waves develop into shock waves.
The explanation of the umbral flash as upward propagating shocks have been confirmed by some observations \citep[e.g.][]{kentischer95, yoon95, brynildsen99, brynildsen03}. 
\citet{lites84} also suggested the possibility that umbral shocks contribute to the heating of the umbral chromosphere.

Some observations have provided evidence for temperature enhancements associated with the umbral flashes \citep[e.g.][]{shibasaki01, iwai17}.
\citet{rodriguez13} derived an enhancement of 1000 K from chromospheric observations in \ion{Ca}{2} 8542 \AA\,obtained using a narrow-band Fabry-P${\rm \acute{e}}$rot based imaging spectrograph.
Using similar observations, \citet{joshi18} reported that the temperature at the chromosphere increases from 3,700 K up to 6,200 K.
\citet{grant18} also measured significant temperature enhancements up to a maximum of $\sim20$ \% through \ion{Ca}{2} 8542 \AA\,observations obtained by another filter-based spectropolarimeter.
Using the \ion{He}{1} 10830 \AA\,triplet, \citet{houston18} obtained temperature fluctuations of $\pm 10$ \% from observations with a grating-based spectrograph.

The heating energy flux required to maintain the umbral chromosphere has been estimated to be $2.6 \times 10^6$ erg cm$^{-2}$ s$^{-1}$ based on the net radiative cooling rate for a semi-empirical sunspot model \citep{avrett81}.
To evaluate the role of shock heating, some authors have investigated energy transported by acoustic waves.
Upward propagating acoustic waves will eventually develop shocks due to the decrease of the density with height, and consequently, their energy will dissipate via shock heating \citep{schwarzschild48}.
%\citet{lites82} estimated Lites 1982 don't measure a decrease of energy flux
%\citet{beck09} quiet region
\citet{felipe11} reproduced observed chromospheric wave signatures through a three-dimensional magnetohydrodynamic numerical simulation perturbed at the photosphere by the observed photospheric velocity fluctuations, and they obtained the acoustic energy flux as a function of the height.
However, the average value of the derived acoustic energy was insufficient to sustain the umbral atmosphere.
Even in the upper photosphere, low acoustic energy fluxes were estimated from Doppler velocity measurements \citep{giovanelli78, kneer81, chae17}.

In contrast to the above measurements, some recent observations have suggested sufficient enough acoustic energy damping in the chromosphere.
\citet{kanoh16} obtained an upward acoustic energy flux of $2.0 \times 10^7$ erg cm$^{-2}$ s$^{-1}$ in the photosphere and $8.3 \times 10^4$ erg cm$^{-2}$ s$^{-1}$ within the lower transition region from Doppler velocity measurements and density estimations obtained using the 
%Solar Optical Telescope \citep[SOT, ][]{tsuneta08, suematsu08, ichimoto08, shimizu08, lites13} onboard 
Hinode \citep{kosugi07} and the Interface Region Imaging Spectrograph \citep[IRIS, ][]{pontieu14}.
In agreement with \citet{kanoh16}, \citet{prasad17} reported a decrease with height of the energy flux at multiple atmospheric heights from intensity oscillations in multiple spectral lines.
% through the Rapid Oscillations in the Solar Atmosphere \citep[ROSA, ][]{jess10} imager and the Hydrogen-Alpha Rapid Dynamics camera \citep[HARDcam, ][]{jess12}.   
Furthermore, \citet{grant15} interpreted observed intensity and area fluctuations as upward propagating sausage-mode oscillations and estimated their energy flux, which decreases with height similar to \citet{prasad17}.
%However, it is not clear whether all of the \textcolor{red}{damped} acoustic energy \textcolor{red}{is transfered} to the shock heating energy or not.

It is currently not clear if the observed reduction of acoustic flux with height implies wave dissipation and heating via shocks, or if it represents loss of wave energy to some other form, such as distortions of the magnetic field, energy transfer to some other wave modes that have less compression, or cascade to unresolved spatial or temporal scales \citep[e.g.][]{cally08, reardon08}.
In order to assess the ability of shocks to heat the umbral chromosphere, we propose to model the thermodynamic and radiative properties of the observed shock plasma.
Our approach to calculate shock heating rates uses spectroscopic observation, while \citet{lee85} uses empirical atmospheric models.

Here, we determine measurements of the shock heating energy rate using spectroscopic observations of the \ion{He}{1} 10830 \AA\, triplet obtained with a fiber-optic-based Integral Field Unit (IFU).
Our analysis allows us to calculate the contribution of the shocks to the umbral chromospheric heating.
To account for the fact that dynamic flashes rapidly change the shape of the spectral line profiles, we take advantage of IFU-fed diffraction-grating-based spectrograph that enables us to measure spectral profiles in one exposure, i.e. without scanning in wavelength as done by Fabry-Perot based instruments, or scanning in space as done by traditional slit-spectrographs.
The IFU also allows us to observe the entire sunspot efficiently with a short enough cadence to resolve the shock dynamics, and it enables us to compare the spatial distribution of shock properties with that of heating signatures.
This IFU technique for such high cadence imaging spectroscopy represents a new class of solar instruments based on multiplexed diffraction-grating-based spectrographs \citep{jaeggli10, lin14, schad17, jurcak19}.
Below we describe details regarding the observations (Section \ref{sec.obs}), the determination of the Mach number and the temperature upstream of the shock (Section \ref{sec.method}), and the results of the shock heating energy rate derived from the shock characteristics (Section \ref{sec.res}).
We then discuss the contributions of the shocks to the umbral chromospheric heating (Section \ref{sec.discussion}) and finally our conclusions (Section \ref{sec.sum}).

%%%%%%%%%%%%%%%%%%%%%%%%%%%%%%%%%%%%%%%%%%%%%
%%%%%%%%%%%%%%%%%%%%%%%%%%%%%%%%%%%%%%%%%%%%%
\section{Observations and data processing}
\label{sec.obs}

%%%%%%%%%%%%%%%%%%%%%%%%%%%%%%%%%%%%%%%%
%                      FIGURE 1 
%%%%%%%%%%%%%%%%%%%%%%%%%%%%%%%%%%%%%%%%
\begin{figure*}
\begin{center}
\includegraphics[angle=0,scale=1.0,width=160mm]{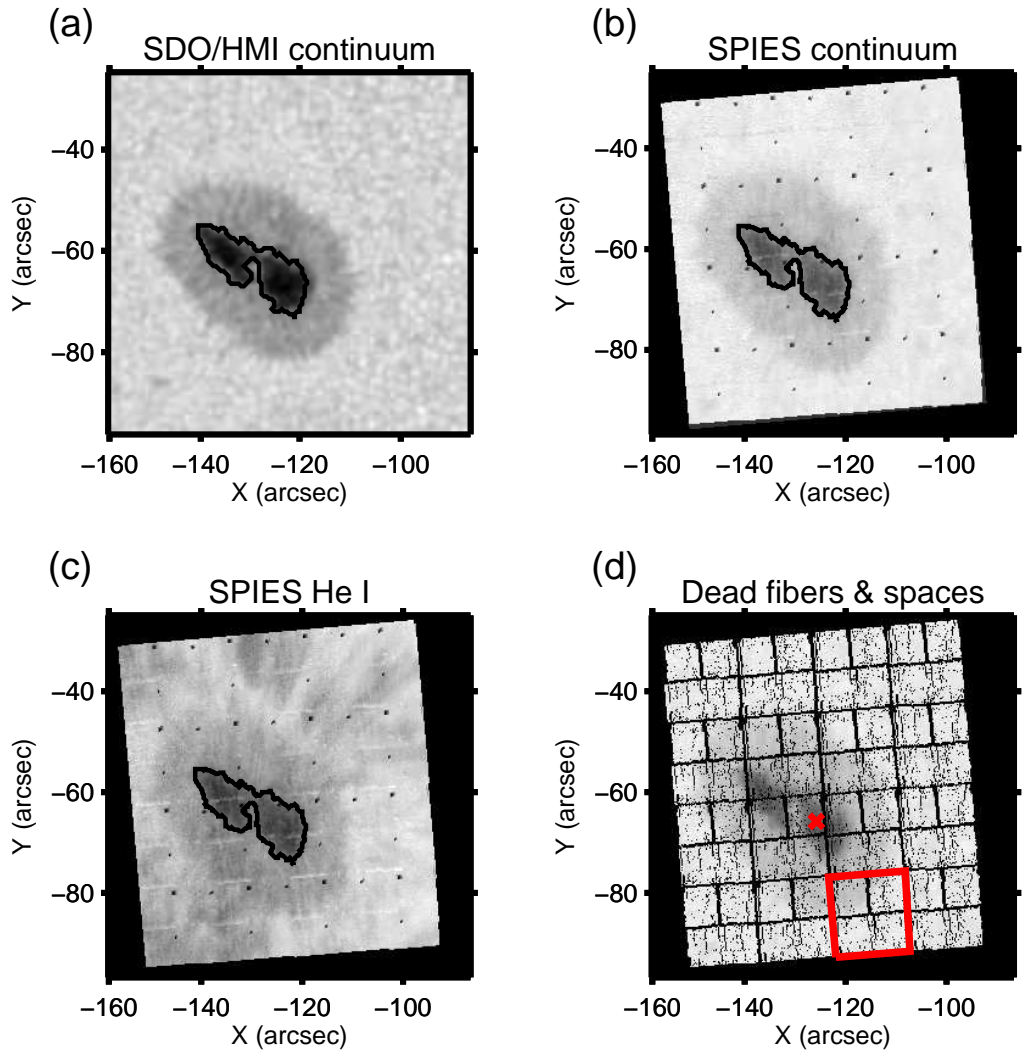}
\end{center}
\caption{
Sunspot images of (a) 617.3 nm continuum intensity taken by SDO/HMI, (b) 1083 nm continuum intensity obtained with SPIES, and (c) line core intensity of \ion{He}{1} 10830 \AA\,acquired with the SPIES at 22:55 UT on 2014 December 7.
The black contour marks the region that we analyze.
(d) Continuum image colored black at the places of dead fibers or gaps between fibers where the instrument is not sensitive in IFU of the SPIES.
The red box indicates single field-of-view of the IFU, and the red cross points where profiles in the figure \ref{fig.2} and \ref{fig.7} was observed. 
		}
\label{fig.1}
\end{figure*}

The leading sunspot in NOAA active region 12227 was observed in a near infrared spectral window containing the \ion{He}{1} 10830 \AA\,triplet with the SpectroPolarimetric Imager for the Energetic Sun \citep[SPIES, ][]{lin06, lin12, schad14} on the NSO's 76 cm Dunn Solar Telescope \citep[DST, ][]{dunn69} between 22:37 and 23:12 UT on 2014 December 7.
The DST high-order adaptive optics system achieved real-time seeing correction and image stabilization \citep{rimmele04}.
The \ion{He}{1} triplet, which is formed in the upper chromosphere, provides a clear signature of shock waves via the temporal change of the Doppler shift \citep{lites86, centeno06}.

SPIES is a prototype instrument for the Diffraction Limited Near Infrared SpectroPolarimeter (DL-NIRSP, https://www.nso.edu/  \\ telescopes/dkist/instruments/dl-nirsp/) of the National Science Fundation's Daniel K. Inouye Solar Telescope \citep[DKIST, ][]{rimmele08, rimmele15, tritschler16}.
Both SPIES and DL-NIRSP employ a fiber-optic-based integral field unit to obtain dispersed spectra within a contiguous two dimensional spatial field of view in a single exposure.
SPIES's IFU \citep{schad14} contains 15,360 birefringent rectangular fibers, and reconfigures a near-square array at a focus of the DST into four slit arrays at the entrance of the near Littrow spectrograph, which are dispersed and imaged simultaneously on a $2048 \times 2048$ infrared detector.
Each individual fiber core has an aspect ratio of 4:1, and thus four adjacent fiber cores are used to form one square spatial sampling pixel.
SPIES therefore allows us to observe a field of 64 $\times$ 60 spatial pixels, corresponding to $15 \times 16$ arcsec$^2$ simultaneously with a high spatial resolution of 0.48 arcsec and a high spectral resolution of 48 m\AA.
The instrumental line width is derived as $29$ m\AA\, from comparisons of observed telluric line profiles and those of a very high resolution solar flux atlas \citep{kurucz84}.
In addition, a larger field of view can be mosaiced by steering the optical field with a two-axis field steering mirror.
In our observations below, we perform a $4 \times 4$ mosaic for a total field of view of $59 \times 65$ arcsec$^2$ and a temporal cadence of 14 seconds. 

After the subtraction by a dark frame, the SPIES dispersed spectral images were reduced via division by a flat field made as follows.
Frames for the flat field were acquired by observing quiet regions around disk center while the telescope conducted a random small amplitude scanning pattern to smear out structures. 
The flat field data were subtracted by a dark frame and averaged.
Next, the averaged frame was divided by another flat field obtained with a lamp, which we inserted in the optical light path to calibrate the transmission profile of a blocking filter.
Moreover, the transmission profile was calibrated by comparing spectral profile in continuum with that of an atlas profile \citep{kurucz84}.
Finally, we applied two-dimensional principal component analysis to the flat frame to remove the solar or telluric spectral features, in analogy to removal of fringes \citep{casini12}.

To facilitate the reconstruction of the spatial image of the IFU, the relative position of each pixel in the detector plane to the location in the IFU entrance plane is determined by placing an undersized square field stop in an upstream location conjugate to the IFU entrance plane.
Orthogonal scans of this field stop using the field steering mirror can then be used to pinpoint the location of each pixel in the field-of-view \citep[see, e.g. ][]{lin06}. 
Figure \ref{fig.1} shows images of the sunspot reconstructed from the SPIES observations as well as that acquired by the Helioseismic and Magnetic Imager \citep[HMI, ][]{scherrer12, schou12} onboard the Solar Dynamics Observatory (SDO).
The sunspot has a light bridge \citep{bray64, lites04, katsukawa07}.
Chromospheric features, for example superpenumbral fibrils \citep{schad13, schad15}, appear in the SPIES \ion{He}{1} 10830 \AA\,image.
The IFU has some dead fibers and gaps between fibers where the instrument is not sensitive (Figure \ref{fig.1} d).
In Figure \ref{fig.1} (b) and (c), the values at the places of the dead fibers and spaces are interpolated by the Delaunay triangulation.
We do not use the interpolated values to derive any quantitative results in the sections below.

%%%%%%%%%%%%%%%%%%%%%%%%%%%%%%%%%%%%%%%%%%%%%
%%%%%%%%%%%%%%%%%%%%%%%%%%%%%%%%%%%%%%%%%%%%%
\section{Diagnosis of shock properties}
\label{sec.method}

To obtain an observationally based estimate for the shock heating energy, one in principle needs to remotely sense the properties of the shocks themselves.
The umbral flashes are clearly slow shocks \citep{centeno06} and the magnetic pressure is larger than the gas pressure in the umbral atmosphere \citep{mathew04}.
%chromosphere is a low beta plasma and far from the equipartition critical point (citation).
In that case, MHD jump conditions (ratios of the state variables) reduce to the hydrodynamic jump for parallel propagating shocks \citep{goedbloed10}. 
%Because some comparisons between observation and theory imply that the shocks propagate along magnetic field lines in the umbral chromosphere \citep[e.g. ][]{centeno06}, we can treat the magnetohydrodynamic shock of the umbral flash as a hydrodynamic shock \citep{goedbloed10}.

We use two different models \citep{weymann60, schatzman49} for the thermodynamic cycle of a parcel of chromospheric plasma that experiences a shock.
In the model of \citet{weymann60}, the shocked plasma radiatively cools at constant specific volume until its entropy matches that of the un-shocked gas, after which it adiabatically expands to the un-shocked pressure and specific volume.
Following  \citet{bray74} \S6.5.3, the heat energy, $q$, dissipated by a hydrodynamic shock per unit mass of gas in this situation can be expressed in terms of the upstream and downstream properties of the plasma as
\begin{equation}
q = \frac{1}{\gamma -1} \frac{R}{\mu} \, T_{{\rm u}} \frac{\rho_{{\rm u}}}{\rho_{{\rm d}}}
      \Bigl[   \frac{p_{{\rm d}}}{p_{{\rm u}}}  - \Bigl(   \frac{\rho_{{\rm d}}}{\rho_{{\rm u}}}  \Bigr)^{\gamma}     \Bigr],
\label{eq.energy_1}
\end{equation}
where $\gamma$ is the polytropic index (the ratio of the specific heats), $R$ is the gas constant, $\mu$ is the mean molecular weight, $T$ is the temperature, $\rho$ is the density, $p$ is the pressure, and the u and d subscripts denote upstream and downstream components of the shock.
Introducing the Mach number of the upstream component, $M \equiv V_{{\rm u}} / \sqrt{\gamma R T_{{\rm u}}/\mu}$, the hydrodynamic jump conditions are derived from the Rankine-Hugoniot relations as
\begin{equation}
\frac{\rho_{{\rm d}}}{\rho_{{\rm u}}} = \frac{V_{{\rm u}}}{V_{{\rm d}}} = \frac{  (\gamma+1) M^2 }{ (\gamma - 1)M^2 +2 },
\label{eq.rh_1}
\end{equation}
\begin{equation}
\frac{p_{{\rm d}}}{p_{{\rm u}}} = \frac{  2 \gamma M^2 -\gamma + 1 }{ \gamma + 1 },
\label{eq.rh_2}
\end{equation}
\begin{equation}
\frac{T_{{\rm d}}}{T_{{\rm u}}} = \frac{ [ 2 \gamma M^2 -\gamma + 1 ] [ (\gamma - 1) M^2 + 2 ] }{ (\gamma + 1)^2 M^2 },
\label{eq.rh_3}
\end{equation}
where $V_{{\rm u}}$ and $V_{{\rm d}}$ are velocities measured in the reference frame co-moving with the shock \citep{landau59}.
In the other model \citep{schatzman49} where the gas is heated and compressed due to the shock, expands adiabatically, and radiates back to the original state, \citet{bray74} give the $q$ as
\begin{equation}
q = \frac{\gamma}{\gamma -1} \frac{R}{\mu} \, T_{{\rm u}} 
      \frac{\rho_{{\rm u}}}{\rho_{{\rm d}}}
      \Bigl[  
                \Bigl(    \frac{p_{{\rm d}}}{p_{{\rm u}}}   \Bigr)^{\frac{1}{\gamma}} -  \frac{\rho_{{\rm d}}}{\rho_{{\rm u}}}     \Bigr].
\label{eq.energy_2}
\end{equation}
Substituting equations (\ref{eq.rh_1}) and (\ref{eq.rh_2}) into equations (\ref{eq.energy_1}) and (\ref{eq.energy_2}) respectively, we are able to derive the shock heating energy per unit mass of gas for either model based only on the Mach number and the temperature upstream of the shock.

In this section, we describe a method to determine the Mach number and the temperature from the observed spectral profiles in the \ion{He}{1} 10830 \AA\,triplet.
First, the Doppler shift and line width were inferred by fitting the spectra with theoretical profiles computed with a radiative transfer equation using an atmospheric model based on a single constant property slab, i.e. a homogeneous plasma (Section \ref{sec.method.single}).
We demonstrate using this single slab model that there are phase relations between the measured Doppler shift and the line width.
As such relations can be interpreted as contributions of the upstream and downstream components of a shock to a line profile, we then fit the profiles with a model consisting of two constant-property slabs for which the temperatures and the macroscopic velocities are constrained by the Mach number as in Equations (\ref{eq.rh_1}) and (\ref{eq.rh_3}) (Section \ref{sec.method.double}).

\citet{hector00, hector00a, hector01} and \citet{centeno05} also analyzed umbral spectra in chromospheric spectral lines using two-component models.
Our method differs from these earlier approaches by taking into account restrictions of the Rankine-Hugoniot relations for state variables of the two components.
%On the other hand, \citet{narukage02} estimated the Mach number and the shock propagation speed of a coronal disturbance associated with a flare using soft X-ray intensity.
In addition to the Mach number, we derive the shock heating energy to assess the ability of shocks to heat the umbral chromosphere.

%%%%%%%%%%%%%%%%%%%%%%%%%%%%%%%%%%%%%%%%%%%%%
%%%%%%%%%%%%%%%%%%%%%%%%%%%%%%%%%%%%%%%%%%%%%
\subsection{Single Slab model}
\label{sec.method.single}

%%%%%%%%%%%%%%%%%%%%%%%%%%%%%%%%%%%%%%%%
%                      FIGURE 2
%%%%%%%%%%%%%%%%%%%%%%%%%%%%%%%%%%%%%%%%
\begin{figure}
\begin{center}
\includegraphics[angle=0,scale=1.0,width=80mm]{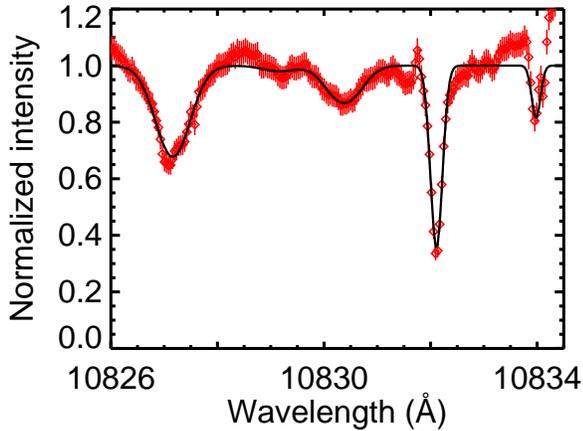}
\end{center}
\caption{
Spectra of the umbra obtained with the SPIES (red diamonds), and fitting results (black solid line) of \ion{Si}{1} 10827 \AA\,with a Voigt function, two telluric lines with two Gaussian functions, and the \ion{He}{1} 10830 \AA\,with a single-slab-model fitting.
The inverted source function, Doppler shift, and line width of the \ion{He}{1} line are $0.63 \pm 0.03$, $0.07 \pm 0.04$ \AA, $0.44 \pm 0.05$ \AA, respectively.
		}
\label{fig.2}
\end{figure}

In order to infer temporal variations of the Doppler shift and the line width of the \ion{He}{1} triplet, we first fit the spectral profiles obtained with SPIES considering a single component model; an example of the fitting results is shown in Figure \ref{fig.2}.
As the linear dispersion and wavelength reference point vary from fiber to fiber, we determine a wavelength calibration for each spectrum by fitting the telluric lines (10832.11 \AA\, and 10833.98 \AA) with two Gaussian profiles.
We fit the \ion{Si}{1} 10827 \AA\,profile with a Voigt function to reduce the effect of its line wing on the fitting for the \ion{He}{1} 10830 \AA\,triplet.

Intensity, $I$, of the theoretical profile of the \ion{He}{1} triplet is obtained by solving a radiative transfer equation using an atmospheric model based on a constant property slab as, 
\begin{equation}
I (\lambda) = S (1-e^{-\tau_{\lambda}}) + I_{{\rm C}} \,e^{-\tau_{\lambda}},
\label{eq.rt_single}
\end{equation}
where $\lambda$ is the wavelength, $S$ is the source function of the slab, and $I_{{\rm C}}$ is the intensity in the continuum.
Optical thickness of the \ion{He}{1} triplet as a function of the wavelength, $\tau_{\lambda}$, can be written as 
\begin{equation}
\tau_{\lambda} = \tau \sum_{i=1}^{3} \omega_{i} \exp{ \Bigl\{ - \Bigl(\frac{\lambda - \lambda_{i} -\lambda_{{\rm D}}}{\Delta \lambda} \Bigr)^2 \Bigr\}},
\end{equation}
where $\tau$ is the optical thickness of the slab at the spectral line center, $\omega_{i}$ are the statistical weights derived from the number of the magnetic sub-levels of the excited state under a constraint of $\sum_{i=1}^{3} \omega_{i}= 1$, $\lambda_{i}$ are the center wavelengths of the triplet lines, $\lambda_{{\rm D}}$ is the Doppler shift,  and $\Delta \lambda$ is the line width.
Even though the constant property slab model is simple, it is a suitable model for reproducing spectral profiles of the \ion{He}{1} 10830 \AA\,line \citep{asensio08}.

Here we fix $I_{{\rm C}}$ to 1 as we normalize each intensity profile by $I_{{\rm C}}$ prior to the fitting.
The fitting cannot determine the optical depth simultaneously with the source function in the optically thin case.
Because the \ion{He}{1} 10830 \AA\,triplet is generally optically thin in the solar spectrum \citep{fontenla93}, we therefore fixed $\tau$ to be 0.5 \citep[see figure 6 of][]{schad15}.
%The \ion{He}{1} 10830 \AA\,triplet is generally optically thin in the solar spectrum \citep{fontenla93}.
%In the optically thin case, the fitting cannot determine $\tau$ simultaneously with $S$, because equation (\ref{eq.rt_single}) is rewritten as $I (\lambda) \approx (S - 1) \tau_{\lambda} + 1$.
%If $\tau$ was small, $S$ would be negative in some fitting results. 
%For example, if $\tau = 0.1$ and $I (\lambda) = 0.7$, $S$ would be $-2$. 
%We therefore selected 0.5 from a range between 0 and 1 as a fixed $\tau$ so that $S$ is positive value.
The quality of the single component fits are satisfactory, although there are systematic residuals.
The residuals at 10831.5 \AA\,and around the telluric line at 10833.98 \AA\, are likely a result of the limited quality of the flat-field calibration.

%%%%%%%%%%%%%%%%%%%%%%%%%%%%%%%%%%%%%%%%
%                      FIGURE 3
%%%%%%%%%%%%%%%%%%%%%%%%%%%%%%%%%%%%%%%%
\begin{figure}
\begin{center}
\includegraphics[angle=0,scale=1.0,width=80mm]{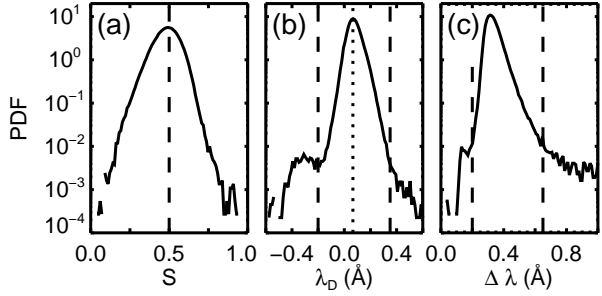}
\end{center}
\caption{
Probability density functions (PDFs) of (a) the source function, (b) the Doppler shift, and (c) the line width inferred with the fitting using the single-slab model.
The dashed vertical lines indicate typical value of the source function, range of the Doppler shift and the line width used in the numerical test for the double-slab-model fitting in Section \ref{sec.discussion}.
The dotted vertical line marks typical value of the Doppler shift (0.067 \AA).
		}
\label{fig.3_1}
\end{figure}

The umbra surrounded by the black solid contours in Figure \ref{fig.1} encompasses roughly 2450 pixels.
Each pixel within this area was fitted at each of the 150 time steps resulting in a total of $\sim 3.7 \times 10^5$ spectral fits.
The results are summarized in Figure \ref{fig.3_1} which shows the distribution of the fitted \ion{He}{1} (a) source function, (b) Doppler shift, and (c) line width.
We do not take into account the values at the dead fibers or spaces to make the histograms.
Since the optical depth is fixed, the source function at the peak of the histogram is determined, and it is approximately equal to 0.5.
Most ($\sim 99.8 \%$) of the Doppler shifts and the line widths vary within the range $-0.2$ \AA\,$< \lambda_{{\rm D}} < 0.35$ \AA\,and $0.2$ \AA\,$ < \Delta \lambda < 0.65$ \AA, respectively.
Since the mode of the Doppler shifts is not equal to zero, the sunspot may be receding from the observer with a speed of $1.85$ km s$^{-1}$ due to the revolution of the Earth and the spins of the Earth and the Sun, although these contributions are not calibrated in our analysis.

%%%%%%%%%%%%%%%%%%%%%%%%%%%%%%%%%%%%%%%%
%                      FIGURE 4
%%%%%%%%%%%%%%%%%%%%%%%%%%%%%%%%%%%%%%%%
\begin{figure}
\begin{center}
\includegraphics[angle=0,scale=1.0,width=80mm]{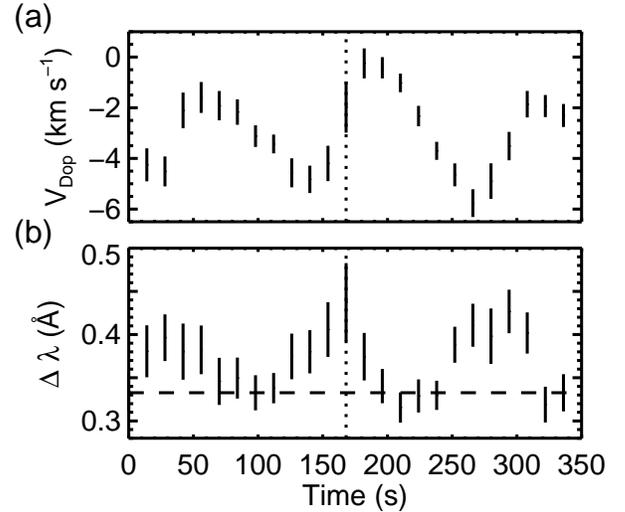}
\end{center}
\caption{
Temporal evolution of the fitted (a) Doppler velocity and (b) line width of the \ion{He}{1} triplet from the same solar location as the profile in Figure \ref{fig.2} and \ref{fig.7} (e.g., for a location corotating with the Sun).
The velocity is positive if the plasma moves up toward the earth.
The vertical dotted lines indicate the time when those profiles were obtained.
For the shock peak at $t = 170$ s, the horizontal dashed line marks the estimate of the pre-shock line width, and hence the upstream temperature, from the immediately preceding minima at $t = 100$ s.
		}
\label{fig.3}
\end{figure}
%%%%%%%%%%%%%%%%%%%%%%%%%%%%%%%%%%%%%%%%
%                      FIGURE 5
%%%%%%%%%%%%%%%%%%%%%%%%%%%%%%%%%%%%%%%%
\begin{figure}
\begin{center}
\includegraphics[angle=0,scale=1.0,width=80mm]{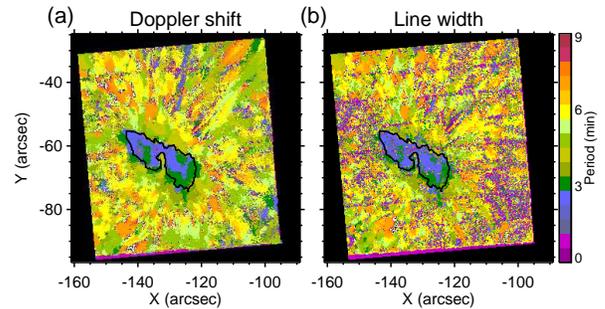}
\end{center}
\caption{
Spatial distributions of the dominant oscillation period in (a) the Doppler shift and (b) the line width.
		}
\label{fig.4}
\end{figure}

Figure \ref{fig.3} shows temporal variations of the inferred Doppler velocity, $V_{{\rm Dop}} \equiv - c \lambda_{{\rm D}}/\lambda_{0}$, and the line width of the \ion{He}{1} triplet for a part of the entire time series at a position with heliocentric coordinates of (S66$^{\circ}$, E126$^{\circ}$) at 22:55UT, where $c$ is the light speed and $\lambda_{0} = 10830$ \AA\,is the wavelength at the line center.  %20141207_umbraosci_077.pro 
When $V_{{\rm Dop}}$ is larger than $-1.85$ km s$^{-1}$, the plasma may be traveling locally upward.
Although the sunspot observed here has a penumbra, the amplitude of the Doppler shift is similar to that of the pore reported in \citet{centeno09}.
Figure \ref{fig.4} displays the spatial variations of the dominant oscillation periods in the Doppler shift and the line width.
Three minute oscillations dominate the umbral chromosphere as many works have confirmed \citep[e.g. ][]{tziotziou07, hector09, reznikova12}.
%\textcolor{red}{Moreover, the dominant period across the umbra is nearly homogeneous, which implies that the inclination angle of the magnetic field with respect to the solar vertical is also homogeneous \citep{reznikova12, yuan12}.}  

%%%%%%%%%%%%%%%%%%%%%%%%%%%%%%%%%%%%%%%%
%                      FIGURE 6
%%%%%%%%%%%%%%%%%%%%%%%%%%%%%%%%%%%%%%%%
\begin{figure}
\begin{center}
\includegraphics[angle=0,scale=1.0,width=80mm]{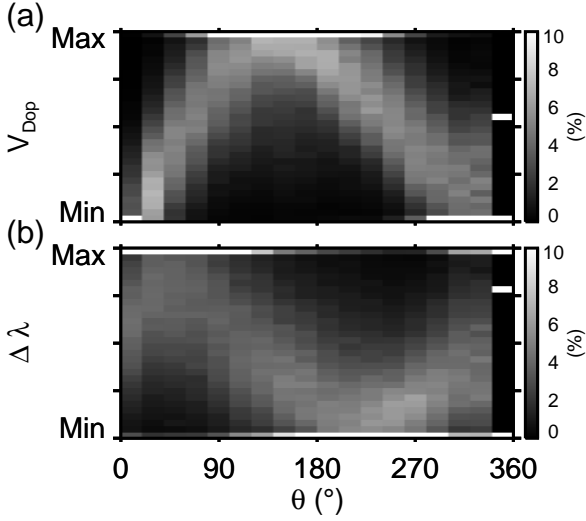}
\end{center}
\caption{
Two dimensional histograms of the Doppler velocity and line width as a function of the phase, which is defined as a time interval normalized by each period for the individual oscillation.
The velocity increases if the plasma is accelerated upward.
The phase is zero when the Doppler velocity has its minimum value.
		}
\label{fig.5}
\end{figure}
%%%%%%%%%%%%%%%%%%%%%%%%%%%%%%%%%%%%%%%%
%                      FIGURE 7
%%%%%%%%%%%%%%%%%%%%%%%%%%%%%%%%%%%%%%%%
\begin{figure}
\begin{center}
\includegraphics[angle=0,scale=1.0,width=80mm]{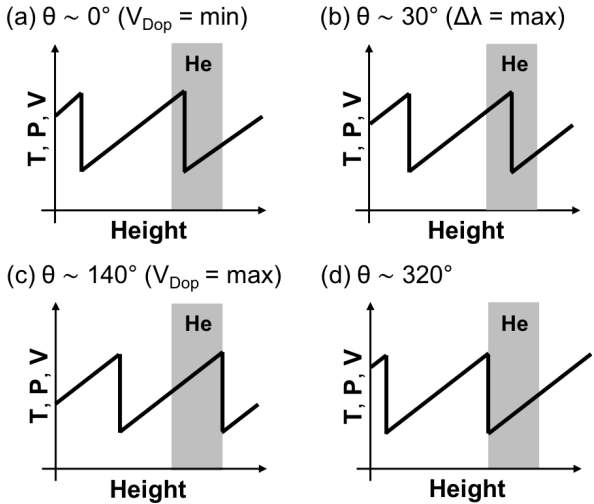}
\end{center}
\caption{
Schematic plots of the temperature, $T$, the pressure, $P$, and the velocity, $V$, as a function of height to interpret the phase relations between the Doppler velocity and the line width.
The gray regions indicate the formation layer of the \ion{He}{1} triplet.
		}
\label{fig.5_1}
\end{figure}

In order to examine phase relations between the Doppler velocity and the line width statistically, we construct two dimensional histograms of their temporal variations vs. the phase as displayed in Figure \ref{fig.5}.
We used values of the umbral data enclosed within the contour in the Figure \ref{fig.1} excluding those at the dead fibers or spaces.
First, we extracted their temporal variations between the time of a minimum Doppler velocity and the next following minimum.
We define the oscillation period as the time interval of the minimum Doppler velocities, and the phase as the time starting from the extracted temporal variation normalized by the period to 360$^{\circ}$. 
To show only phase relations, the measured amplitude is normalized by the peak-to-peak value of the extracted time series and is stacked to make the two dimensional histograms.
Each histogram at each phase space bin is normalized by the total number.

The temporal variation in the $V_{{\rm Dop}}$ reveals a sawtooth pattern, that is, the phase of the maximum $V_{{\rm Dop}}$ is smaller than the half period.
In addition, the line width has its maximum value during the time $V_{{\rm Dop}}$ is increasing.
The sawtooth pattern and the enhancement of the line width during the $V_{{\rm Dop}}$ increases indicate that a shock wave was upwardly propagating through the formation layer of the \ion{He}{1} triplet during this phase.
We interpret the increase of the line width as the superposition of newly shocked rising plasma and falling plasma from the passage of the previous shock.
Both components contribute to the formation of the line profile during the phase, as \citet{tian14} demonstrated for an IRIS observation in the \ion{Si}{4} line formed at the transition region.

Figure \ref{fig.5_1} shows schematic plots to explain this interpretation.
Because the sunspot was located near disk center (Figure \ref{fig.1}), the Doppler velocity corresponds to a velocity component normal to the solar surface.
The sawtooth shocks propagate upward in the umbral chromosphere.
About $\theta \sim 30^{\circ}$, the shock should be about the middle of the formation layer of the \ion{He}{1} triplet, because the shocked rising plasma and the falling plasma equally contribute to form the broadest line in this period.
For $\theta \sim 140^{\circ}$, the Doppler velocity reaches its maximum, because the falling upstream component exits the line-forming region and stops to contribute the line formation. 
%In $\theta > 140^{\circ}$, the line width is decreasing with the temperature of the downstream component. 
For $\theta \sim 320^{\circ}$, the line starts to broaden as the next shock reaches the \ion{He}{1} formation layer.  
%However, there is not typical phase, in which the line starts to broaden, since the histograms about $\theta \sim 320^{\circ}$ are broad.

%%%%%%%%%%%%%%%%%%%%%%%%%%%%%%%%%%%%%%%%%%%%%
%%%%%%%%%%%%%%%%%%%%%%%%%%%%%%%%%%%%%%%%%%%%%
\subsection{Double Slab model with the Rankine-Hugoniot relations}
\label{sec.method.double}

Since the temporal variations in the Doppler velocity and the line width indicate that both the upstream and downstream components of the shock should contribute to the line formation, we now repeat the fits to the \ion{He}{1} spectral profiles of the umbra with theoretical profiles using an atmospheric model based on two constant property slabs, of which temperatures and macroscopic velocities are constrained by equations (\ref{eq.rh_1}) and (\ref{eq.rh_3}).
We limit our two-slab fitting analysis to the line profiles from the time during each oscillation when the width reaches its maximum, that is, the contribution to the line intensities from the plasma on each side of the shock is maximized, in the umbra surrounded by the black contour in the Figure \ref{fig.1}.

Here, we assume that 
(1) the shock propagates parallel to the magnetic field.
It is reasonable because the umbral flash is suggested to be a slow magnetic hydrodynamic shock \citep{centeno06} and reduces the magneto-hydrodynamic shock to a hydrodynamic shock.
Since the sunspot was located near disk center (Figure \ref{fig.1}), (2) the shocks are also assumed to propagate along the line-of-sight.
These assumptions greatly simplify the analysis.
In addition, we assume that (3) the broadening of the triplet absorption profiles are due only to the thermal and the instrumental effects.  %\textcolor{blue}{, which is reasonable because ?.}
Effects of the assumption of the zero non-thermal line broadening on the results are discussed in Section \ref{sec.discussion}.

Modeled profiles for the \ion{Si}{1} line and the telluric lines remain as before, whereas the two-slab radiative transfer equation for the upstream and downstream components of the shock, which propagates in the formation layer of the \ion{He}{1} triplet, is now written as:
\begin{eqnarray}
I (\lambda) = S_{{\rm u}} (1-e^{-\tau_{{\rm u}}(\lambda)}) &+& 
                     S_{{\rm d}} e^{-\tau_{{\rm u}}(\lambda)} (1-e^{-\tau_{{\rm d}}(\lambda)})  \nonumber \\
                     &+&  I_{{\rm C}} \,e^{-\tau_{{\rm u}}(\lambda) - \tau_{{\rm d}}(\lambda)},
\end{eqnarray}
\begin{equation}
\tau_{{\rm u}}(\lambda) = \tau_{{\rm u}0} \sum_{i=1}^{3} \omega_{i} \exp{ \Bigl\{ - \Bigl(\frac{\lambda - \lambda_{i} -\lambda_{{\rm Du}}}  { \Delta \lambda_{{\rm u}} } \Bigr)^2 \Bigr\}},
\end{equation}
\begin{equation}
\tau_{{\rm d}}(\lambda) = \tau_{{\rm d}0} \sum_{i=1}^{3} \omega_{i} \exp{ \Bigl\{ - \Bigl(\frac{\lambda - \lambda_{i} -\lambda_{{\rm Dd}}}  { \Delta \lambda_{{\rm d}} } \Bigr)^2 \Bigr\}},
\end{equation}
where u and d indicate upstream and downstream components, respectively.
Thermal broadening is proportional to the square root of the temperature.
Here we do not consider unresolved motions and assume the observed line widths and thermal line width, $\Delta \lambda_{{\rm th}}$, are related by $\Delta \lambda_{s}^2 = \Delta \lambda_{{\rm I}}^2 +  \Delta \lambda_{{\rm th, s}}^2$, where $s$ indicates the upstream or downstream plasma.
Therefore, using equation (\ref{eq.rh_3}),
%From the upstream line width, $\Delta \lambda_{{\rm u}}$, and the Mach number of the upstream component, $M$, the line width of the downstream component, $\Delta \lambda_{{\rm d}}$, can be \textcolor{red}{derived using equation (\ref{eq.rh_3})} as  
\begin{eqnarray}
\Delta \lambda_{{\rm d}} = \Biggl[ \Delta &\lambda_{{\rm th, u}}^2&    \frac{(2\gamma M^2 - \gamma+1) [(\gamma-1)M^2 +2] }{(\gamma+1)^2 M^2}    \nonumber   \\
&+& \Delta \lambda_{{\rm I}}^2    \Biggr]^{0.5},
\label{eq.derive_mach}
\end{eqnarray}
where $\Delta \lambda_{{\rm I}} = 29$ m\AA\,is the instrumental line width.
Assuming a monoatomic ideal gas, we fix $\gamma$ to $5/3$.
The line width and the Mach number of the upstream component also determine velocities in the shock frame, $V_{{\rm u}}$ and $V_{{\rm d}}$, of the upstream and downstream components as
\begin{equation}
V_{{\rm u}} = M \biggl\{  \frac{ \gamma R m c^2 }{2 \mu k_{{\rm B}} \lambda_{0}^{2}} (\Delta \lambda_{{\rm u}}^2 - \Delta \lambda_{{\rm I}}^2)      \biggr\}^{0.5},
\end{equation}
\begin{equation}
V_{{\rm d}} = V_{{\rm u}} \frac{(\gamma-1)M^2 +2}{(\gamma+1)M^2},
\end{equation}
where $m$ is the mass of helium and $k_{{\rm B}}$ is the Boltzmann constant.
We derived the mean molecular weight, $\mu$, as $1.29$ from the solar atomic abundance \citep{asplund09}, and umbral atmospheric models \citep{maltby86} at the \ion{He}{1} triplet formation height \citep{felipe10}.  
%mean molecular weight mu=0.61 Mariska 1993
Denoting propagation speed of the shock, $U$, the Doppler shift of both components are calculated from velocities in the observer reference frame as 
\begin{equation}
\lambda_{{\rm Du}} = \lambda_{0} (V_{{\rm u}} +U) /c,
\label{eq.14}
\end{equation}
 and 
\begin{equation}
\lambda_{{\rm Dd}} = \lambda_{0} (V_{{\rm d}} +U) /c.
\label{eq.15}
\end{equation}
In summary, the parameters of the radiative transfer equation are $M$, $U$, $\Delta \lambda_{{\rm u}}$, $S_{{\rm u}}$, $S_{{\rm d}}$, $ \tau_{{\rm u}0}$, $ \tau_{{\rm d}0}$, $I_{{\rm C}}$, and $\Delta \lambda_{{\rm I}}$.

As for the single slab model fitting (Sec. \ref{sec.method.single}), the optical depths ($\tau_{{\rm u}0}$ and $\tau_{{\rm d}0}$) and the intensity in the continuum, $I_{{\rm C}}$, are fixed to be 0.5 and 1, respectively.   
In addition, we assume the upstream line width, $\Delta \lambda_{{\rm u}}$, to be the minimum value of the line width found from the single slab model fitting just from the previous cycle at each spatial position, that is, each fitting uses a different fixed value for $\Delta \lambda_{{\rm u}}$.
The horizontal dashed line in the Figure \ref{fig.3} marks the fixed value of the upstream line width to fit a profile obtained at the time indicated by the vertical dotted line.
This approach can be justified by the fact the shock does not affect the temperature upstream in the chromosphere.
Therefore, the variable parameters in the fit are only $M$, $U$, $S_{{\rm u}}$, and $S_{{\rm d}}$.
Before doing the fitting, we subtracted time averaged residual profiles of the single slab model fitting from the observed spectra to avoid fitting the residuals at 10831.5 \AA\, with one of double slabs (Figure \ref{fig.2}).

%%%%%%%%%%%%%%%%%%%%%%%%%%%%%%%%%%%%%%%%
%                      FIGURE 8
%%%%%%%%%%%%%%%%%%%%%%%%%%%%%%%%%%%%%%%%
\begin{figure}
\begin{center}
\includegraphics[angle=0,scale=1.0,width=80mm]{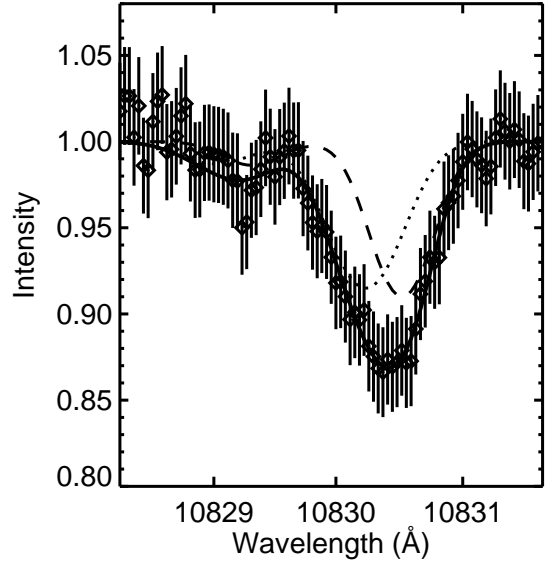}
\end{center}
\caption{
Spectral profile of the umbra in the \ion{He}{1} 10830 \AA\, (black diamonds), and fitting result using double-slab model (black solid line).
The spectra is a part of that shown in the Figure \ref{fig.2}.
The black dashed and dotted line indicate individual contributions of the double slab.
The inverted source function of upstream and downstream components, the Mach number and the shock speed in the upstream are, respectively, $0.75 \pm 0.08$, $0.76 \pm 0.10$, $1.43 \pm 0.12$, and $-15 \pm 2$ km s$^{-1}$.
		}
\label{fig.7}
\end{figure}
 %%%%%%%%%%%%%%%%%%%%%%%%%%%%%%%%%%%%%%%%
%                      FIGURE 9
%%%%%%%%%%%%%%%%%%%%%%%%%%%%%%%%%%%%%%%%
\begin{figure}
\begin{center}
\includegraphics[angle=0,scale=1.0,width=80mm]{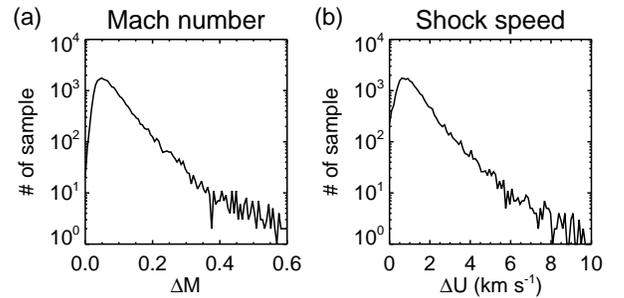}
\end{center}
\caption{
Histograms of (a) the error in the Mach number, and (b) the error in the shock speed.
		}
\label{fig.error}
\end{figure}

An example of the fittings with the double-slab model is shown in Figure \ref{fig.7}.
The separate slab contributions, as shown by the black dashed and dotted lines, are calculated by solving Equation (\ref{eq.rt_single}) using parameters of each slab.
The line width of the redder component is fixed by using the estimate described above, for the case indicated by the horizontal dashed line in the Figure \ref{fig.3}(b), which is the value of the line width at time $t=100$ s, the minimum immediately preceeding the shock peak at time $t=170$ s.
The fitting of profiles with double slabs gives us the ratio of the line widths between the upstream and downstream components and their Doppler velocities.
From the line-width ratio, we are able to determine the Mach number in the upstream component using equation (\ref{eq.derive_mach}).
%Moreover, the equation (\ref{eq.rh_1}) allows us to derive the velocity ratio in the shock frame from the Mach number.
%Finally, comparing the velocity ratio in the observer frame with that in the shock frame provides us the propagation speed of the shock. 
Substituting the definition of the Mach number into equation (\ref{eq.14}), the shock propagation speed is given as  
\begin{equation}
U = \frac{\lambda_{{\rm Du}}}{\lambda_0} c - M \sqrt{\frac{\gamma R T_{{\rm u}}}{\mu}}.
\label{eq.16}
\end{equation}
Because $T_{\rm u}$ is equal to $\frac{1}{2} m c^2 k_{{\rm B}}^{-1} \lambda_{0}^{-2} (\Delta \lambda_{{\rm u}}^2 - \Delta \lambda_{{\rm I}}^2)$, the shock speed is derived from the inferred Mach number, the Doppler shift of the upstream component, and the fixed upstream line width.

Our initial values used in the fitting are $M=1.001$, $U=c \lambda_{{\rm Du}} /  \lambda_{0} - M \sqrt{\gamma R T_{\rm u} / \mu}$, $S_{{\rm u}}=S$, and $S_{{\rm d}}=0.98$, where $S$ is derived from the single slab model fitting. 
%, and $T_{\rm u}$ is derived from the fixed upstream line width, $\Delta \lambda_{\rm u}$ as $\frac{1}{2} m c^2 k_{{\rm B}}^{-1} \lambda_{0}^{-2} (\Delta \lambda_{{\rm u}}^2 - \Delta \lambda_{{\rm I}}^2)$.

We performed multiple non-linear least squares fits using a gradient-expansion algorithm by randomly selecting the $\lambda_{{\rm Du}}$, which determines the initial estimate of the shock speed, within a range $\lambda_{{\rm D}} < \lambda_{{\rm Du}} < \lambda_{{\rm D}} + \Delta \lambda$, until the number of attempts to fit a profile reaches to 1,000 or the number of success fits reaches 50 where success is defined both by a converged least-squares solution and that the two source functions are both less than 0.9, because the source function of the \ion{He}{1} triplet is generally smaller than 0.9 when the optical depth is fixed to be 0.5 (Figure \ref{fig.3_1}).
%If a source function is greater than 0.9, the line depth formed by the slab is too shallow to contribute the line profile.  
We obtained solutions from 19,717 profiles, of which 19,125 profiles provide us 50 successfull fittings with changing the initial estimate of the shock speed, $U$. %20190215_1.pro We applied the method to 25,309 profiles, and 
Here, the fitting solution for a line profile is defined as a parameter set of a fitting that gives us a median value in the Mach number.

The error is defined as the larger of the two values of standard deviation given by the individual solution and the outputs of the successfully randomized fittings.
Figure \ref{fig.error} shows histograms of errors of the 19,717 profiles.
Peaks of the histograms are at 0.05 in the error of the Mach number and at 1 km s$^{-1}$ of the shock speed.

%%%%%%%%%%%%%%%%%%%%%%%%%%%%%%%%%%%%%%%%%%%%%
%%%%%%%%%%%%%%%%%%%%%%%%%%%%%%%%%%%%%%%%%%%%%
\section{Results}
\label{sec.res}

%%%%%%%%%%%%%%%%%%%%%%%%%%%%%%%%%%%%%%%%
%                      FIGURE 10
%%%%%%%%%%%%%%%%%%%%%%%%%%%%%%%%%%%%%%%%
\begin{figure*}
\begin{center}
\includegraphics[angle=0,scale=1.0,width=160mm]{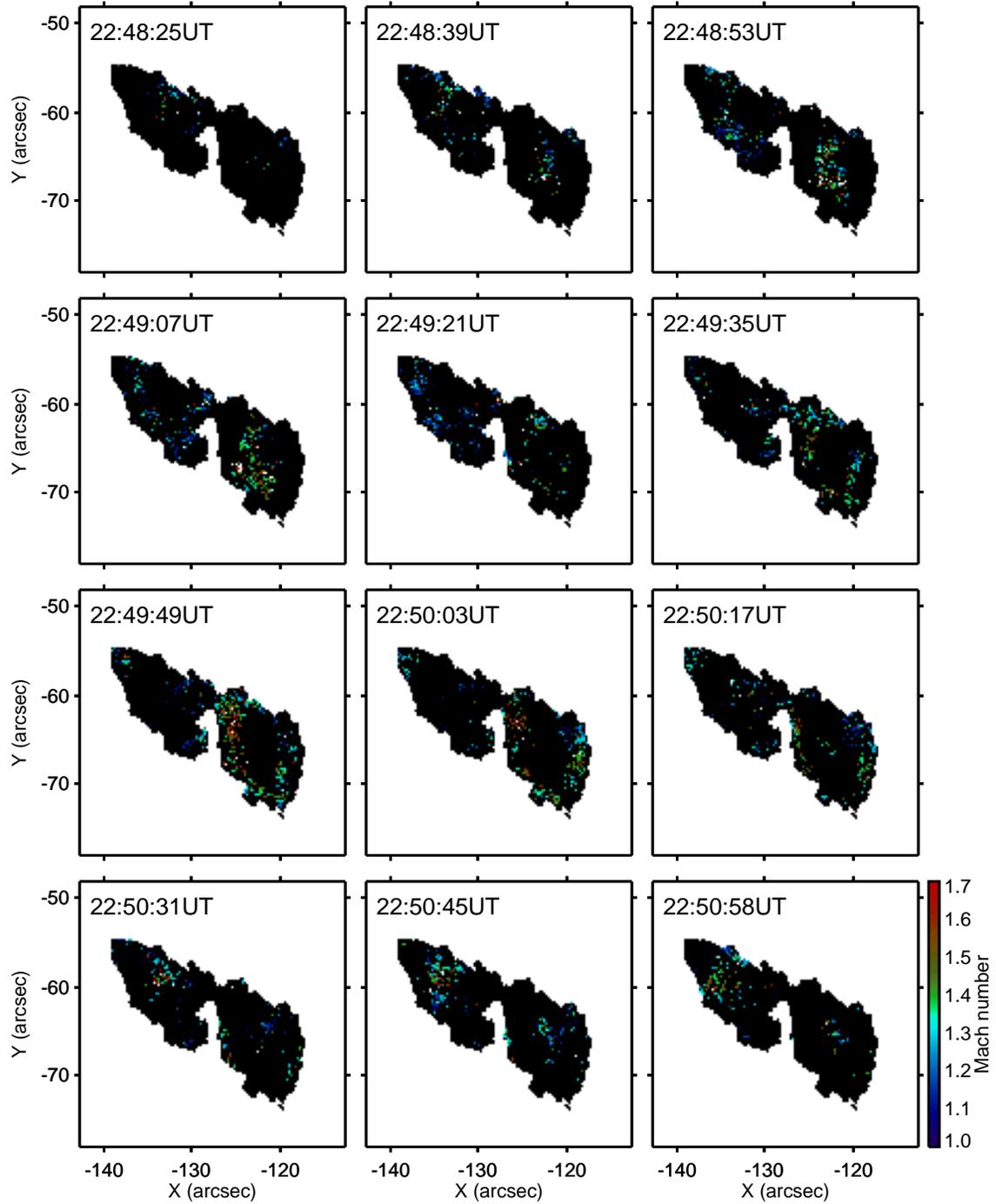}  %normal
\end{center}
\caption{
Spatial and temporal variations of the upstream Mach number across the sunspot umbra.
A movie is available online.% (file name is 20141207$\_$umbraosci$\_$133.gif).
%渦は今回の論文では扱わないから泣く泣く止める。
		}
\label{fig.8}
\end{figure*}

We use the SPIES instrument to determine the properties of shocks associated with umbral flashes.
Spectral observations taken with each fiber allow us to directly determine the properties of each shock, while the spatial coverage of the IFU allows us to study variations in the spatial and temporal characteristics of the shocks.
Figure \ref{fig.8} shows the spatial and temporal variations of the Mach number over the umbra.
For this figure, the Mach number at the place of the dead fibers or spaces in the umbra (Figure \ref{fig.1}d) is determined from a profile expected by a triangulation.
The two dimensional plus temporal data show expanding phase fronts that originate near center of the right umbra, and expanded quasi-circularly to the edge of the umbra similar to what has been reported for umbral flashes \citep[e.g. ][]{rouppe03, nagashima07}.
In the figure \ref{fig.8} animation, the evolution of the shocks show spiral patterns \citep{ariste13, ariste16, su16, jess17, felipe19, kang19}.
The radial propagation of the pattern may be apparent propagation rather than physical wave propagation \citep{bogdan06}.

%%%%%%%%%%%%%%%%%%%%%%%%%%%%%%%%%%%%%%%%
%                      FIGURE 11
%%%%%%%%%%%%%%%%%%%%%%%%%%%%%%%%%%%%%%%%
\begin{figure}
\begin{center}
\includegraphics[angle=0,scale=1.0,width=80mm]{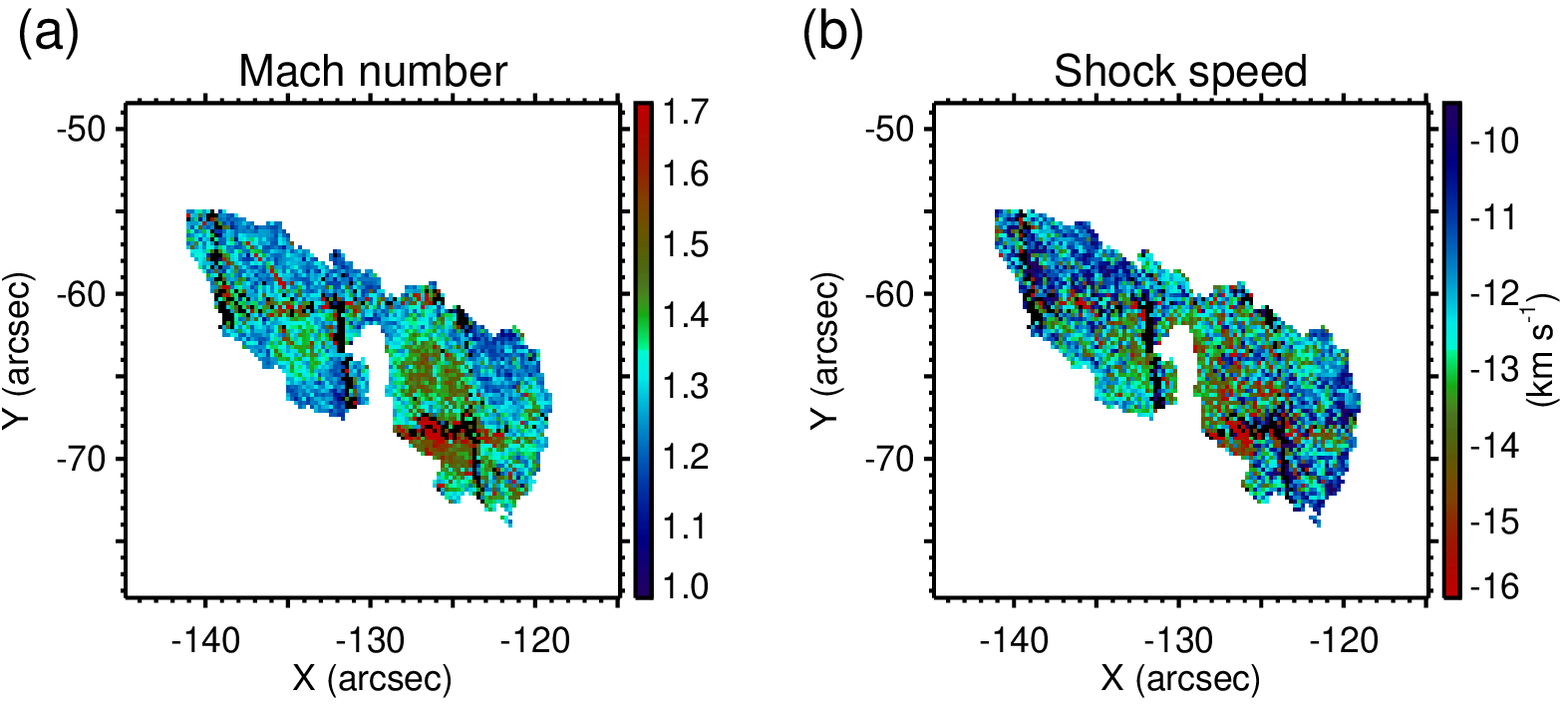}
\end{center}
\caption{
Spatial distributions of (a) the Mach number and (b) the Shock speed averaged at each spatial point. 
%mean(mach[mach ge 1])=1.35
		}
\label{fig.10}
\end{figure}

Spatial distributions of the average Mach number and average shock speed at each spatial point are displayed in Figure \ref{fig.10}.
The average values are derived without the values at the dead fibers or spaces.
Because the sunspot image moves across the IFU field during the observation, we are able to derive the average values across almost the entire umbra.
For the shock speed, negative value means upflow, since the definition of the sign is the same as that of the Doppler shift (opposite to that of the Doppler velocity $V_{{\rm Dop}}$).
Roughly speaking, higher Mach numbers are associated with faster shocks.
However, the spatial distribution of the Mach number is slightly different from that of the shock speed.
The light bridge is the near gap between two lobes of the sunspot.
The Mach number seems to be high about the center of each lobe of the sunspot.
On the other hand, the shock speed seems to be largest near the light bridge and gradually decreases away from the light bridge.

%%%%%%%%%%%%%%%%%%%%%%%%%%%%%%%%%%%%%%%%
%                      FIGURE 12
%%%%%%%%%%%%%%%%%%%%%%%%%%%%%%%%%%%%%%%%
\begin{figure}
\begin{center}
\includegraphics[angle=0,scale=1.0,width=80mm]{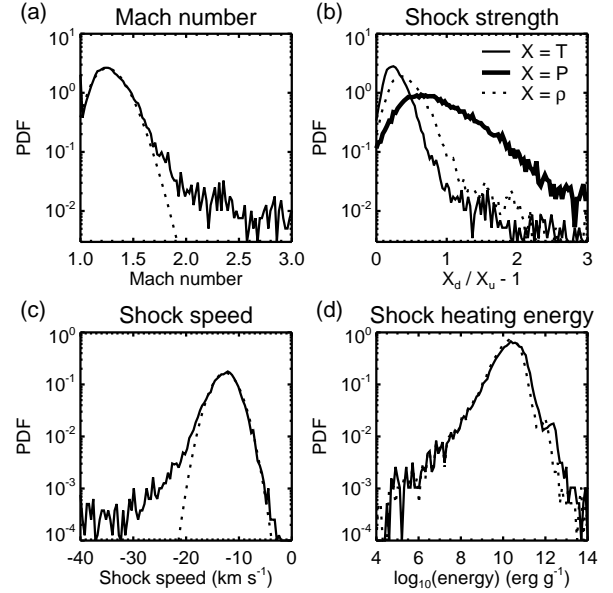}
\end{center}
\caption{
Probability density functions (PDFs) of (a) the Mach number, (b) the temperature (thin solid line), pressure (thick solid line) and density (dotted line) enhancements, (c) the shock speed, and (d) the shock heating energy. 
The dotted lines in (a) is a fitted logarithmic normal distribution and in (c) is a fitted normal distributions.
The solid and dotted lines in (d) are calculated with the equations (\ref{eq.energy_1}) and (\ref{eq.energy_2}), respectively.
%expected value of the Mach = 1.30
		}
\label{fig.9}
\end{figure}

Figure \ref{fig.9} plots the probability density functions (PDFs) of the (a) Mach number, (b) temperature enhancement, pressure enhancement, density enhancement at the shock front, the (c) shock speed, and the (d) shock heating energy per unit mass of plasma. 
The PDFs are derived without the values at the dead fibers and spaces.
The temperature, pressure, and density enhancements at the shock front are calculated from the Mach number using equations (\ref{eq.rh_1}), (\ref{eq.rh_2}) and (\ref{eq.rh_3}).

The temperature enhancement at the peak of the PDF is approximately equal to 0.24, which is consistent with those reported by \citet{grant18} and \citet{houston18}.
The Mach number is 1.24, and the shock speed is $-12$ km s$^{-1}$ at the peak of the PDFs. 
The PDF of the Mach number is fitted with a logarithmic normal distribution $2.60 \exp{\{-0.5 [ (\ln{|M|} - 0.226)/0.116 ]^2 \} }$, and the PDF of the shock speed is fitted with a normal distribution $0.168 \exp{\{-0.5 [ (U + 12.62)/2.29 ]^2 \} }$.
We derive the shock heating energy per unit mass of plasma with the two models of thermodynamic cycles discussed in Section \ref{sec.method} using the equations (\ref{eq.energy_1}) and (\ref{eq.energy_2}) (Figure \ref{fig.9} d).
Their PDFs are almost identical, and they have peaks at $2.3\times10^{10}$ erg ${\rm g^{-1}}$ and $1.8\times10^{10}$ erg ${\rm g^{-1}}$.
Their expected values ($\int q\, {\rm PDF} (q) \, dq$) are $1.8\times10^{10}$ erg ${\rm g^{-1}}$ and $1.4\times10^{10}$ erg ${\rm g^{-1}}$.
We conclude typical shock heating energy per unit mass of plasma is approximately equal to $2 \times10^{10}$ erg ${\rm g^{-1}}$.
% total(dx*x*y)/total(dx*y)=1.4 or 1.8 x 10^10 erg/g

%%%%%%%%%%%%%%%%%%%%%%%%%%%%%%%%%%%%%%%%
%                      FIGURE 13
%%%%%%%%%%%%%%%%%%%%%%%%%%%%%%%%%%%%%%%%
\begin{figure}
\begin{center}
\includegraphics[angle=0,scale=1.0,width=80mm]{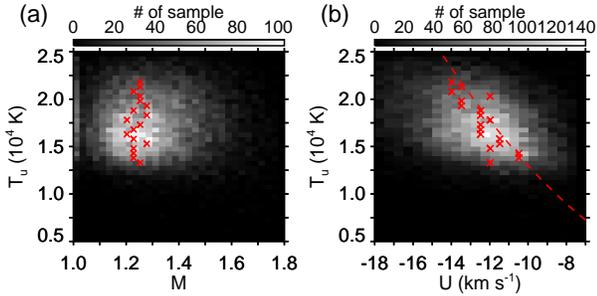}
\end{center}
\caption{
(a) Temperature in the upstream of the shock vs. Mach number in the upstream, and (b) vs. shock speed.
The red crosses denote maximum densities at each temperature bin in a range between $13,000$ K and $22,000$ K.
The red dashed line shows the sound speed derived from the upstream temperature.
		}
\label{fig.11}
\end{figure}

Figure \ref{fig.11} shows two-dimensional histograms of the temperature upstream of the shocks vs. the Mach number (\ref{fig.11}a) and the shock speed (\ref{fig.11}a).
In Figure \ref{fig.11}a, the crosses denote the most likely Mach number for each upstream temperature histogram bin.
We find no relation between $T_{{\rm u}}$ and $M$ (linear Pearson correlation coefficient of -0.02).
Conversely, we do find a weak linear relation in \ref{fig.11}b, with greater upstream temperatures having greater shock speeds (Pearson coefficient -0.32).
The red dashed line marks the temperature dependent sound speed ($C_{s}=\sqrt{\gamma R T_{{\rm u}}/\mu}$), 
with an offset of $1.85$ km s$^{-1}$ to take into account the the relative velocity between the sunspot and the observers (Section \ref{sec.method.single}).
Because the offset of the shock speed for the maximum density at each temperature bin (red crosses) is equal to $-0.5 \pm 0.6$ km s$^{-1}$, we conclude that the saw-tooth shocks propagate upward at the sound speed: they are in the weak shock regime.

%%%%%%%%%%%%%%%%%%%%%%%%%%%%%%%%%%%%%%%%%%%%%
%%%%%%%%%%%%%%%%%%%%%%%%%%%%%%%%%%%%%%%%%%%%%
\section{Discussion}
\label{sec.discussion}

We use spectra of the \ion{He}{1} 10830 \AA\,triplet from the SPIES instrument to determine the properties of shocks that pass through the umbral chromosphere.
From the measured Mach number of the shock and the temperature of the upstream plasma we found shock heating, $q$, of $\sim 2 \times10^{10}$ erg ${\rm g^{-1}}$ per shock cycle, independent of the assumed thermodynamic shock cycle.
Moreover, we find that the shocks propagate with the sound speed, and the Mach number does not depend on the temperature.

Does the shock heating energy balance the radiative energy losses in the umbral chromosphere?
Since the formation height of the \ion{He}{1} 10830 \AA\,triplet is estimated by \cite{felipe10} as 1038 - 1208 km from a height where the continuum optical depth at 500 nm is unity, the net radiative cooling rate at the formation layer can be $0.02 - 0.1$ erg ${\rm cm^{-3}}$ ${\rm s^{-1}}$ \citep{avrett81}.
%Assuming the density, $\rho$, as $2.5 \times 10^{-11}$ g cm$^{-3}$ at the formation layer according to an umbral model of \cite{maltby86}, and a typical shock period of $\tau=180$ from Figure \ref{fig.4}, we determine the shock heating rate as $\rho \epsilon / \tau = 3 \times 10^{-3}$ erg ${\rm cm^{-3}}$ ${\rm s^{-1}}$, which is 1 - 10 \% of the required amount of energy to compensate the radiative energy losses in the umbral chromosphere.
Assuming the density, $\rho$, as $1 \times 10^{-11}$ g cm$^{-3}$ at the formation layer according to an umbral flash model of \cite{bose19}, and a typical shock period of $\tau=180$ s from Figure \ref{fig.4}, we determine the shock heating rate as $\rho q / \tau = 1 \times 10^{-3}$ erg ${\rm cm^{-3}}$ ${\rm s^{-1}}$, which is 1 - 5 \% of the required amount of energy to compensate the radiative energy losses in the umbral chromosphere.
This conclusion is consistent with that of a calculation of the shock heating rates for two umbral atmospheric models \citep{lee85}.
%The optical depths of the upstream and downstream components were fixed to be $0.5$ in the fitting (Section \ref{sec.method.double}).
We performed the same fittings as that described in section \ref{sec.method.double} but using different values of the optical depths from 0.5 ($\tau_{{\rm u}0} = \tau_{{\rm d}0}= 0.2$ and $\tau_{{\rm u}0} = \tau_{{\rm d}0}= 0.8$).
Their results do not change the conclusion that the shock heating rate is insufficient to maintain the umbral chromosphere.

\begin{figure}
\begin{center}
\includegraphics[angle=0,scale=1.0,width=80mm]{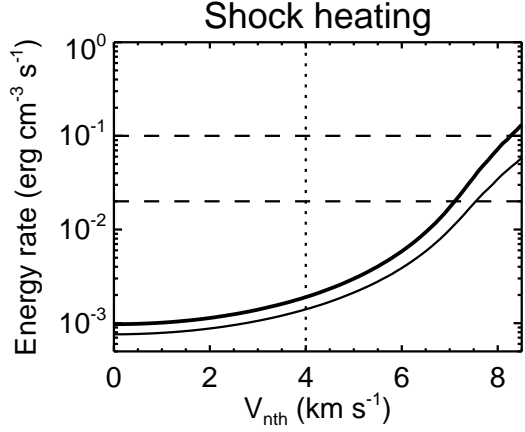}
\end{center}
\caption{
	Expected values of the shock heating energy rate as a function of the non-thermal velocity ($V_{nth}$).
	The thick and thin solid lines indicate the expected values estimated with the two models of thermodynamic cycles.
	The horizontal dashed lines show the range of the net radiative cooling rate at the formation layer of the \ion{He}{1} triplet \citep{avrett81}.
	The vertical dotted line marks non-thermal velocity at the formation layer for an umbral flash model \citep{bose19}.
		}
\label{fig.nth}
\end{figure}

We neglected the contribution of non-thermal motions to the spectral line broadening.
However, \cite{bose19} derived the non-thermal velocity to be $\sim 4$ km s$^{-1}$ at the formation layer of the \ion{He}{1} triplet for the umbral flash.
Figure \ref{fig.nth} shows expected values of the shock heating energy rate as a function of the non-thermal velocity, $V_{nth}$.
First, we calculated the Mach number, $M'$, solving equation \ref{eq.rh_3} from ratios of redefined temperatures as $T'_{{\rm u}} \equiv T_{{\rm u}} - m V_{nth}^2 / 2 k_{B}$ and $T'_{{\rm d}} \equiv T_{{\rm d}} - m V_{nth}^2 / 2 k_{B}$ with an assumption that $V_{nth}$ upstream and downstream of the shocks are the same.
Next, the shock heating energy per unit mass of plasma for the two models of thermodynamic cycles is derived with equations (\ref{eq.energy_1}) and (\ref{eq.energy_2}) from $M'$ and $T'_{{\rm u}}$, and its expected values are obtained as described in section \ref{sec.res}.
Finally, the plotted shock heating energy rates are estimated as discussed above using the expected values.
Although the heating energy rates increase with the non-thermal velocity, the rates at the non-thermal velocity of 4 km s$^{-1}$ are insufficient to maintain the umbral chromosphere.

%%%%%%%%%%%%%%%%%%%%%%%%%%%%%%%%%%%%%%%%
%                      FIGURE 14
%%%%%%%%%%%%%%%%%%%%%%%%%%%%%%%%%%%%%%%%

\begin{figure}
\begin{center}
\includegraphics[angle=0,scale=1.0,width=80mm]{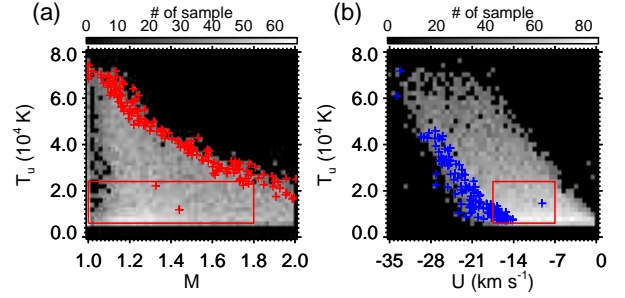}
\end{center}
\caption{
As Figure \ref{fig.11}, but obtained from numerical test runs for 10,000 synthetic spectral profiles.
The red boxes indicate the parameter ranges displayed in the Figure  \ref{fig.11}.
The red and blue crosses are solutions for the synthetic spectral profiles calculated with line widths between $0.64$ \AA\,and $0.65$ \AA, and Doppler velocities between $-0.20$ \AA\,and $-0.19$ \AA, respectively.
		}
\label{fig.13}
\end{figure}

The derived upstream temperature exhibits a significant linear relation with the shock speed but not with the Mach number.
To demonstrate the relations are not systematic results from the fitting, we performed a Monte Carlo simulation by synthesizing and fitting 10,000 spectral profiles calculated with randomly distributed parameters within the measured typical range $-0.2$ \AA\,$ < \lambda_{{\rm D}} < 0.35$ \AA\,and $0.2$ \AA\,$ < \Delta \lambda < 0.65$ \AA\,(Figure \ref{fig.3_1}).
The upstream line width, $\Delta \lambda_{{\rm u}}$, is fixed to be a randomly distributed parameter within a range $0.2\,{\rm \AA} < \Delta \lambda_{{\rm u}} < \Delta \lambda$.%, where $\Delta \lambda$ is the line width derived from the single slab model fitting. 
Approximately 10\% of the synthetic spectra could not be fit successfully according to the criteria described in section \ref{sec.method.double}.
Figure \ref{fig.13} displays the fit parameters of the successfully fitted profiles.
%The upstream temperature is determined from the line width of the upstream component, which we fix \textcolor{red}{during} the fitting to be the measured minimum value before the Doppler-shift change from redshift to blueshift.
%On the other hand, the shock speed is derived from comparing the velocity ratios between upstream and downstream components in the observer frame with that in the shock frame as explained in the section \ref{sec.method.double}.
%The velocities in the observer frame corresponds to the measured wavelengths of the two components, and their ratio in the shock frame is determined from the Mach number, which \textcolor{red}{is} derived from the measured line widths of the two components.
%Therefore, the linear relations among the upstream temperature, the Mach number, and the shock speed results from independent measured values.

%%%%%%%%%%%%%%%%%%%%%%%%%%%%%%%%%%%%%%%%
%                      FIGURE 15
%%%%%%%%%%%%%%%%%%%%%%%%%%%%%%%%%%%%%%%%

To understand how the model varies with each parameter we can consider isoparametric lines by holding one parameter fixed at a time.
For instance, the upper right boundary in Figure \ref{fig.13} (a) corresponds to solutions with fixed line width of the single slab model, which we use for the synthesis: red pluses mark solutions for which $0.64$ \AA\,$< \Delta \lambda <$ $0.65$ \AA.
%For Figure \ref{fig.13} (a), the right downing upper boundary is formed by solutions for the synthetic spectral profiles calculated with line widths between $0.64$ \AA\,and $0.65$ \AA, because larger line width gives higher Mach number, that is, line width ratio between downstream and upstream components, at each upstream temperature bin.
For Figure \ref{fig.13} (b), the lower left edge is formed by solutions for the synthetic spectral profiles calculated with Doppler shifts between $-0.20$ \AA\,and $-0.19$ \AA.
We explain this behavior by noting that the Doppler shift of the upstream component $\lambda_{{\rm Du}}$ is nearly the Doppler shift of the spectral line itself (see the dashed and solid lines in Figure \ref{fig.7}).
Making that approximation in Equation (\ref{eq.16}), we find that 
\begin{equation}
U  \sim \frac{\lambda_{{\rm D}}}{\lambda_0} c - M \sqrt{\frac{\gamma R T_{{\rm u}}}{\mu}}.
\label{eq.17}
\end{equation}
Since the Mach number does not have significant dependence on the upstream temperature (Figure \ref{fig.11} a), the absolute value of the shock speed increases with the upstream temperature for a fixed Doppler shift.
%\textcolor{red}{Therefore, the edges of the distributions are made by the boundaries of the parameter ranges.}

There is no systematic relation within the distribution of the test solutions without boundaries, and their boundaries are determined from the observed parameter ranges.%, are out of the displayed range in the Figure \ref{fig.11}.
\,In addition, the shock speed is roughly determined from the independent parameters, Doppler shift and the upstream temperature (Equation \ref{eq.17}).
Therefore, we conclude that the observed significant relations among the shock speed, the Mach number, and the upstream temperature do not systematically result from the fitting method.

%\textcolor{blue}{{\bf non thermal line width}}

\citet{bogdan03} performed two dimensional MHD simulations of wave propagation in a variety of solar-like magnetized atmosphere.
They introduced waves using sinusoidal photospheric driving.
The wave amplitude of velocity fluctuations normalized by the sound speed increased as $\rho ^{-2/5}$ due to the decrease in density with height in the stratified atmosphere.
Eventually, the waves gave rise to shocks.
They found that slow MHD shocks propagate slightly faster than the sound speed in strong magnetic field concentrations, such as sunspots.
Although it is not significant, our measured shock propagation speed tends to be faster than the sound speed by $0.5$ km s$^{-1}$ (Figure \ref{fig.11}b).

In the \citet{bogdan03} simulations, after the formation of the shocks, the velocity saturates due to the numerical viscosity.
If the viscosity was negligibly small as expected in the real solar atmosphere \citep{vranjes13}, the Mach number could be written as
\begin{equation}
%M \sim 1+ \frac{2}{5} v_{0}   \int_{0}^{H} \Bigl[ \frac{1}{\Lambda C_{s}} {\rm exp} \Bigl( \frac{2h}{5 \Lambda} \Bigr)  \Bigr] {\rm d} h,
M \approx 1 + \frac{v_{0}}{C_{s}} {\rm exp} \Bigl( \frac{2h}{5 \Lambda} \Bigr),
\label{eq.18}
\end{equation}
where $h$ is the height above the photosphere, $v_{0}$ is the amplitude of the velocity fluctuation at the photosphere, $\Lambda$ is the density scale height ($=R T / \mu g$), and $g$ is the gravity acceleration.
If the formation height of the \ion{He}{1} triplet is $1000$ km \citep{felipe10} and $v_{0}/C_{s}=0.05$, then the Mach number at $h=1000$ km is approximately equal to 1.2 for $T=1.25 \times 10^4$ K and 1.1 for $T=2.50 \times 10^4$ K.
Therefore, the Mach number would decrease with increasing temperature.
However, we observed the Mach number is independent of the upstream temperature.
Because $v_{0}$ should not depend on $H$ and $T$, we propose that the Mach number may saturate in the umbral atmosphere due to energy loss of the shocks along a flux tube.
As examples, using magnetohydrodynamic simulations of magnetic flux tubes, \cite{takasao13} shows the slow shock energy should be carried across magnetic field lines by the fast-mode magnetohydrodynamic waves generated at a place where the sound speed is equal to the Alfv${\rm \acute{e}}$n speed, as produced by the shocks, and \cite{shelyag16} shows efficient ambipolar dissipation of Alfv${\rm \acute{e}}$n waves transformed from acoustic shocks.
In addition, if the shock does not propagate parallel to the magnetic field lines and it has not yet reached the steady state, intermediate shock substructures within the shock may dissipate the acoustic energy \citep{snow19}.
%Although their objects are not sunspots, \citep{reardon08} found a signature of turbulence generated from shocks, 

%%%%%%%%%%%%%%%%%%%%%%%%%%%%%%%%%%%%%%%%%%%%%
%%%%%%%%%%%%%%%%%%%%%%%%%%%%%%%%%%%%%%%%%%%%%
\section{Summary}
\label{sec.sum}

We derived the shock heating energy per unit mass of plasma from the spectra in  the near infrared \ion{He}{1} triplet measured with the IFU spectrometer, SPIES, on the Dunn Solar Telescope.
The SPIES, which is a prototype instrument of the DL-NIRSP of the forthcoming DKIST, allows us to compare shock properties with that of heating signatures over  the entire umbra, even though they change rapidly in time. 

In order to determine the shock parameters, we fit the measured spectral profiles in the \ion{He}{1} 10830 \AA\,triplet with theoretical profiles computed with the radiative transfer equation using an atmospheric model based on two constant property slabs with temperatures and macroscopic velocities constrained by the Rankine-Hugoniot relations.
As a result, the typical shock heating energy per unit mass of plasma is $\sim 2 \times10^{10}$ erg ${\rm g^{-1}}$ per shock cycle, which is insufficient to maintain the umbral chromosphere.

A much stronger conclusions are that shock propagation at the sound speed is consistent with them being weak shocks 
%(and in particular, weak, acoustically dominated shocks in a low-beta environment)
, and the Mach number does not depend on the temperature upstream of the shocks.
If the viscosity is negligibly small as expected in the actual solar atmosphere, the Mach number should decrease with the temperature increase.
Therefore, we propose that energy loss of the shocks may suppress the amplification of the Mach number in the umbral atmosphere.

\acknowledgments
We extend thanks to Prof. J. R. Kuhn, Dr. G. I. Dima, Dr. X. Sun, Dr. M. Kramar, and Dr. A. Fehlmann who discussed with us every week.
The authors are also grateful to Dr. H. Lin for providing the SPIES, and Dr. A. Hiller for discussing the interpretation. 
The National Solar Observatory (NSO) is operated by the Association of Universities for Research in Astronomy, Inc. (AURA), under cooperative agreement with the National Science Foundation.
The SDO data are provided courtesy of NASA/SDO and the AIA and HMI science teams.


\vspace{5mm}
\facilities{DST(SPIES)}


\software{Solar Soft}

\bibliography{arxiv}



\end{document}